# Automation Security

**Timur Z. Mirzoev**

*The Department of Electronics and Computer Technology*
*College of Technology, Indiana State University, Terre Haute IN, 47807*

January 2007

**Introduction**

Over the past decade, the traditional corporate network has come under siege from a proliferation of viruses and malicious intruders. As a result, numerous network security technologies, methodologies, and policies have been developed to secure the business system. In comparison, process control systems, with their reliance on proprietary networks and hardware, have long been considered immune to cyber attacks. Now, with the move to open standards such as Ethernet, TCP/IP, and web technologies, the plant floor requires its own security procedures to eliminate undesirable access to production systems and to maximize the uptime and efficiency of process networks[1].

Web-based Automated Process Control systems are a new type of applications that use the Internet to control industrial processes with the access to the real-time data. Supervisory control and data acquisition (SCADA) networks contain computers and applications that perform key functions in providing essential services and commodities (e.g., electricity, natural gas, gasoline, water, waste treatment, transportation) to all Americans. As such, they are part of the nation's critical infrastructure and require protection from a variety of threats that exist in cyber space today. By allowing the collection and analysis of data and control of equipment such as pumps and valves from remote locations, SCADA networks provide great efficiency and are widely used[2]. IEEE Standard 1402-2000, *Guide for Electric Power Substation Physical and Electronic Security*, states:

> "The introduction of computer systems with online access to substation information is significant in that substation relay protection, control, and data collection systems may be exposed to the same vulnerabilities as all other computer systems. As the use of computer equipment within the substation environment increases, the need for security systems to prevent electronic intrusions may become even more important."

---

[1] E. Byres, G. Gillespie, *Plan for Network security*, Industrial Networking, Fall 2002

[2] US Department of Energy, *"21 Steps to Improve Cyber Security of SCADA Networks"*, retrieved from http://www.ea.doe.gov/pdfs/21stepsbooklet.pdf on November 16, 2003

It is fortunate, that vulnerabilities of computer systems are well known and well documented. However, it is important to successfully apply that knowledge to industrial networks. Figure 1. presents electronic intrusion vulnerabilities for industrial networks.

Over the last 20 years, there were increasingly more complex, sophisticated, and capable control systems installed. These have included more sophisticated basic and advanced regulatory control, advanced process control, optimizers, and artificial intelligence[3]. The field of intelligent instrumentation is going through a major transition. The first "smart" transmitters were pioneered by just a few companies some 15 years ago. After a rather slow start in the market, one can say that in the past five years, this technology has finally received acceptance in the plant instrumentation and maintenance communities. The benefits of intelligent devices-enhanced performance, improved

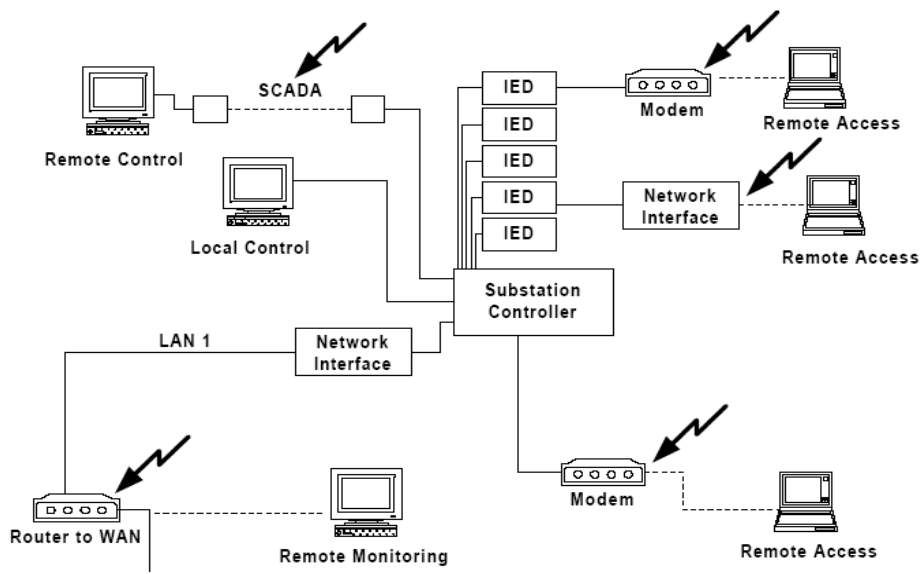

Figure 1. Electronic Intrusion Vulnerabilities

Source: Paul Oman, Edmund O. Schweitzer, III, and Jeff Roberts, "*Safeguarding IEDS, Substations, and SCADA systems against electronic intrusions*", Schweitzer Engineering Laboratories, Inc. 2001

---

[3] William L. (Bill) Mostia (1999). *Out of Control*. Control for the Process Industries, July 1999

reliability, and the ability to provide self-diagnostics have been recognized as contributors to improve the processes they control.  Furthermore, prices have come down significantly, thereby making the purchase of intelligent devices affordable and justifiable[4].

It is well-known that IT security has been addressed many times in the past decade.  Many scary examples of hackers taking over businesses, companies' web servers and etc., indicate the need for continuous improvement of security systems.  Those examples mainly show IT vulnerability in business or economic area of peoples' lives; however, consequences of a hacker intruding a power plant security or compromising a web-based automated process control system running PLCs (Programmable Logic Controllers) and other control devices could result in more dramatic outcomes than just shutting down an e-commerce web server or interrupting an online business.

> In October, an Australian man was sent to prison for two years after he was found guilty of hacking into a waste management system and causing millions of liters of raw sewage to spill out into local parks, rivers, and even the grounds of a Hyatt Regency hotel.  He exacted this revenge after the area's council rejected his job application to work as a control engineer at the treatment plant.  It appears that the problem may be more widespread than most process engineers believe.  The incident database maintained by the British Columbia Institute of Technology Internet Engineering Lab now contains more than a dozen cyber attacks involving process control process systems in all sectors of manufacturing[5].

In order to successfully protect industrial networks, there should be a plan developed for security procedures of industrial networks.  In the next chapters of this paper there will be some suggestions on how to properly deal with industrial security issues; some possible solutions will also be presented.

---

[4] Peter L. Schellekens, *"A Vision for Intelligent Instrumentation in Process Control"*. *Control Engineering Online*, October 2001.

[5] Eric Byres, *"Can't happen at your site?"* www.isa.org, February 2002

**Identifying Threats and Risks**

Web-based control systems are becoming more used with everyday and the online-based network control systems create more possibilities for intruders to invade a network control system. Internet technologies such as RSView32 by Rockwell Automation, LabView by National Instruments[6] are providing great applications for control networks. However, it is necessary to mention that the network and data integrity are not the primary concern of those software packages.

Motivations for electronic attacks against web-based control systems may follow the same patterns seen in attacks on Internet E-commerce sites. They can be categorized into five broad groups[7]:

1. *Hackers* are computer users who access unauthorized systems *simply because they can*. The relatively benign hacker is motivated by curiosity or the challenge of exploration, without overt malicious intent. Others are malicious with the intent of gaining notoriety or causing damage.

2. *Espionage* is the act of gaining industrial or political advantage by information gathering through both legal and illegal means. Much espionage gathers information through publicly available resources such as web pages, product descriptions, and promotional literature. Other espionage activities include insider access, theft, and illegal surveillance to acquire confidential information.

---

[6] Software packages that allow engineers control and view control or device networks via WWW.

[7] Paul Oman, Edmund O. Schweitzer, III, and Jeff Roberts, "*Safeguarding IEDS, Substations, and SCADA systems against electronic intrusions*", Schweitzer Engineering Laboratories, Inc. 2001

3. *Sabotage* is usually rooted in desires for personal, economic, or political gain caused through the destruction of your competitor's assets, organizational structure, and/or market share. "Hactivism" is an emerging form of sabotage where hackers deface or damage corporate Information Technology assets in the name of some radical cause.

4. *Electronic Theft* is the theft of credit and/or personal identity information, frequently stored in corporate IT systems, that can be used in subsequent fraudulent schemes. Losses in the U.S. alone are estimated to be billions of dollars.

5. *Vandalism* is the destruction of property value without personal gain, as distinct from sabotage because it is typically haphazard, random, and relatively localized.

Creation of secure industrial networks is possible only when potential threats and risks are identified. Figure 2. depicts generalization of potential threats.

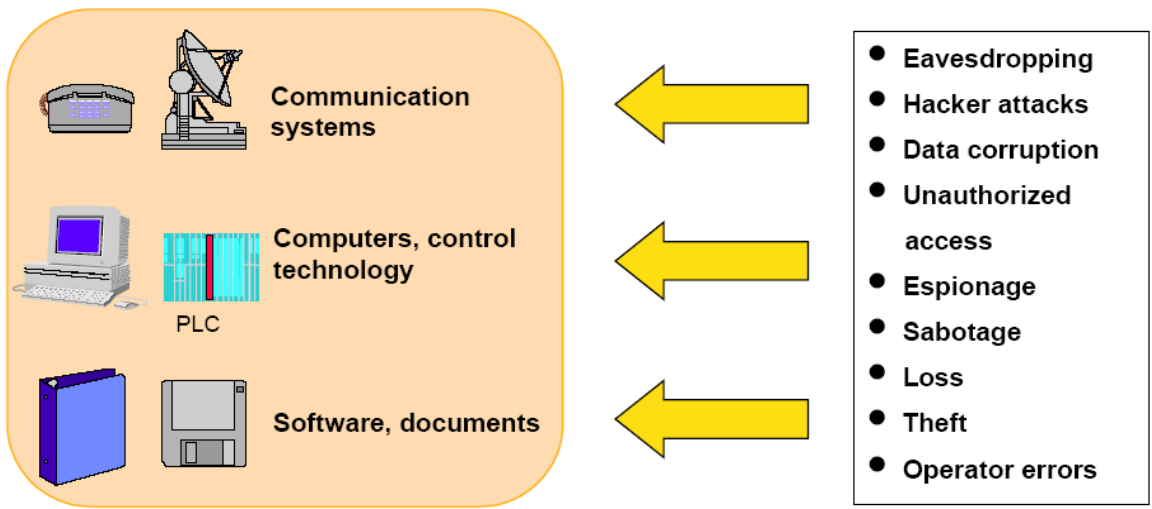

Figure 2. Security Risks and Types of Assault

Source: Siemens, *Information Security in Industrial Communications*, October 1999

The following presents an example of threats identified for electric power systems[8]:

1. The expanded use of public protocols to interconnect protective equipment and SCADA systems (e.g. TCP/IP and UCA over Ethernet LANs/WANs).

2. Increased dial-in and network access to remote sites through public communication services (e.g., public phones, Internet).

3. Instability in the electric power utility job market, creating disgruntled employees and ex-employees, caused by deregulation and mergers.

4. Increased competition for electricity generation and T&D services creating pressure to downsize, streamline, automate and cut costs, also causing disgruntled employees and ex-employees.

5. Instability in the electric power service, caused by deregulation and increased competition, creating disgruntled customers.

6. Increased public access to transmission system data mandated by FERC 888/889.

7. Increased terrorism worldwide and increased foreign government-sponsored terrorism and information warfare targeted against North America.

8. Rapid growth of a computer-literate population and widespread dissemination of hacker tools.

9. Increased electronic theft, recreational hacking, and hacktivism (i.e., the destruction of electronic assets for a political or socioeconomic cause).

Federal Bureau of Investigation also identified some threats of critical infrastructure that are presented in Table 1.

---

[8] Paul Oman, Edmund O. Schweitzer, III, and Jeff Roberts, "*Safeguarding IEDS, Substations, and SCADA systems against electronic intrusions*", Schweitzer Engineering Laboratories, Inc. 2001

Table 1.  Threats to Critical Infrastructure Observed by the FBI

Source: Federal Bureau of Investigation

| Threat | Description |
|---|---|
| Criminal groups | There is an increased use of cyber intrusions by criminal groups who attack systems for purposes of monetary gain. |
| Foreign intelligence services | Foreign intelligence services use cyber tools as part of their information gathering and espionage activities. |
| Hackers | Hackers sometimes crack into networks for the thrill of the challenge or for bragging rights in the hacker community. While remote cracking once required a fair amount of skill or computer knowledge, hackers can now download attack scripts and protocols from the Internet and launch them against victim sites. Thus, while attack tools have become more sophisticated, they have also become easier to use. |
| Hacktivists | Hacktivism refers to politically motivated attacks on publicly accessible Web pages or e-mail servers. These groups and individuals overload e-mail servers and hack into Web sites to send a political message. |
| Information warfare | Several nations are aggressively working to develop information warfare doctrine, programs, and capabilities. Such capabilities enable a single entity to have a significant and serious impact by disrupting the supply, communications, and economic infrastructures that support military power—impacts that, according to the Director of Central Intelligence,a can affect the daily lives of Americans across the country. |
| Insider threat | The disgruntled organization insider is a principal source of computer crimes. Insiders may not need a great deal of knowledge about computer intrusions because their knowledge of a victim system often allows them to gain unrestricted access to cause damage to the system or to steal system data. The insider threat also includes outsourcing vendors. |
| Virus writers | Virus writers are posing an increasingly serious threat. Several destructive computer viruses and "worms" have harmed files and hard drives, including the Melissa Macro Virus, the Explore.Zip worm, the CIH (Chernobyl) Virus, Nimda, and Code Red. |

Resulting consequences of a possible intrusion of industrial or SCADA network could be as follows[9]:

• Shut down the substation or any portion of the subsystem controlled by the compromised device, either immediately or in a delayed manner.

• Change protection device settings to degrade reliability of the device and, subsequently, the electric service provided by the substation.

• Gather control and protection settings information that could be used in a subsequent attack.

---

[9] Paul Oman, Edmund O. Schweitzer, III, and Jeff Roberts, "*Safeguarding IEDS, Substations, and SCADA systems against electronic intrusions*", Schweitzer Engineering Laboratories, Inc. 2001

- Change or perturb the data in such a manner as to degrade electric service or cause loss of service.
- Plant malicious code that could later trigger a delayed or coordinated attack.
- Shut down the regional service controlled by that SCADA system, either immediately or in a delayed manner.
- Steal or alter metering data gathered by the SCADA system.
- Use the SCADA system as a backdoor into the corporate IT system to obtain customer credit and personal identity information commonly used in electronic theft.

**Development of security procedures for industrial control networks**

In order to successfully protect industrial control networks, E. Byres and G. Gillespie suggest protecting control network integrity by starting not from technology, but from creation of an industrial cyber security policy. The authors of the article "Plan for Network Security[10]" present the following characteristics of a typical industrial cyber security policy:

- Establish a secure production communication infrastructure and a stable processing environment.
- Effectively manage the risk of security exposure and intrusion.
- Communicate the responsibilities of users, management, and information system staff for the protection of information and provide direction toward honoring this commitment.
- Provide a requirements outline for those staff charged with technical implementation.
- Promote understanding and compliance with all applicable laws and regulations.

---

[10] E. Byres, G. Gillespie, *Plan for Network security*, Industrial Networking, Fall 2002

- Limit company liability and preserve management's options in the event of a security incident.

The responsibility for an industrial cyber security policy should reside at the highest organizational structure. Below that should be a production-based technical team charged with standards implementation, technical management, and enforcement. The consistent application of the security policy will reduce the risk and liability of all persons involved[11].

US Department of Energy suggest following these 21 steps to improve industrial control network security[12]:

1. Identify all connections to SCADA networks
2. Disconnect unnecessary connections to the SCADA network
3. Evaluate and strengthen the security of any remaining connections to the SCADA network
4. Harden SCADA networks by removing or disabling unnecessary services.
5. Do not rely on proprietary protocols to protect your system.
6. Implement the security features provided by device and system vendors
7. Establish strong controls over any medium that is used as a backdoor into the SCADA network.
8. Implement internal and external intrusion detection systems and establish 24-hour-a-day incident monitoring.
9. Perform technical audits of SCADA devices and networks, and any other connected networks, to identify security concerns.
10. Conduct physical security surveys and assess all remote sites connected to the SCADA network to evaluate their security

---

[11] E. Byres, G. Gillespie, *Plan for Network security*, Industrial Networking, Fall 2002

[12] US Department of Energy, *"21 Steps to Improve Cyber Security of SCADA Networks"*, retrieved from http://www.ea.doe.gov/pdfs/21stepsbooklet.pdf on November 16, 2003

11. Establish SCADA "Red Teams" to identify and evaluate possible attack scenarios.

12. Clearly define cyber security roles, responsibilities, and authorities for managers, system administrators, and users

13. Document network architecture and identify systems that serve critical functions or contain sensitive information that require additional levels of protection.

14. Establish a rigorous, ongoing risk management process.

15. Establish a network protection strategy based on the principle of defense-in-depth.

16. Clearly identify cyber security requirements

17. Establish effective configuration management processes

18. Conduct routine self-assessments

19. Establish system backups and disaster recovery plans.

20. Senior organizational leadership should establish expectations for cyber security performance and hold individuals accountable for their performance

21. Establish policies and conduct training to minimize the likelihood that organizational personnel will inadvertently disclose sensitive information regarding SCADA system design, operations, or security controls.

In comparison to business IT security, it might seem that there are no clear security guidelines developed for SCADA and web-based control networks. However, general suggestions given here could be extremely useful for each particular industry.

**Recommendations and Conclusions**

Jonathan Pollet argues, that SCADA security is often overlooked[13]:

---

[13] Jonathan Pollet, *SCADA Security Strategy*, PlantData Technologies August 8, 2002

Now, more than ever, utilities and oil and gas producers are taking precautions to ensure the security of their information systems, implementing protections for known potential points of access via the Internet. The problem lies in the fact that many SCADA systems involve various components that bridge the world between IT and Facility Engineering. There is typically not a lot of teamwork and collaboration between the IT security efforts and the Data Acquisition teams that are focused on delivering field and facility data reliably to company stockholders and end users. Everyone assumes that it is an IT issue, but IT staffs are rarely have the expertise and background in SCADA and industrial automation systems to adequately develop security tests or measures for these systems.

A survey by Riptech found that about 70 percent of the operating manuals for those systems are available to the public. These systems for the process industries are particularly vulnerable because they were not designed for use with Internet applications.

The White House report on critical infrastructure protection recommends the three following steps to secure network-based infrastructure[14]:

"We suggest consideration of these immediate actions prior to the completion of a formal risk assessment:

(1) Isolate critical control systems from insecure networks by disconnection or adequate firewalls,

(2) Adopt best practices for password control and protection, or install modern authentication mechanisms,

(3) Provide for individual accountability through protected action logs or the equivalent."

---

[14] The White House Office of the Press Secretary, White House Communications on Critical Infrastructure Protection, October 22, 1997: http://www.julieryan.com/Infrastructure/IPdoc.html.

Robert F. Dacey, Director of Information Security Issues of United States General Accounting Office, states the following:

> A significant challenge in effectively securing control systems is the lack of specialized security technologies for these [control networks] systems. As I [Robert F. Dacey] previously mentioned, the computing resources in control systems that are needed to perform security functions tend to be quite limited, making it very difficult to use security technologies within control system networks without severely hindering performance.
>
> Although technologies such as robust firewalls and strong authentication can be employed to better segment control systems from enterprise networks, research and development could help address the application of security technologies to the control systems themselves. Information security organizations have noted that a gap exists between current security technologies and the need for additional research and development to secure control systems
>
> Several steps can be considered when addressing potential threats to control systems, including[15]:
>
> • Researching and developing new security technologies to protect control systems.
>
> • Developing security policies, guidance, and standards for control system security. For example, the use of consensus standards could be considered to encourage industry to invest in stronger security for control systems.
>
> • Increasing security awareness and sharing information about implementing more secure architectures and existing security technologies. For example, a more secure architecture might be attained by segmenting control networks with robust firewalls and strong authentication. Also, organizations may benefit from educating management about the cybersecurity risks related to control systems and sharing successful practices related to working across organizational boundaries.

---

[15] Robert F. Dacey, *Critical Infrastructure Protection*, Testimony Before the Subcommittee on Technology, Information Policy, Intergovernmental Relations, and the Census, House Committee on Government Reform, United States General Accounting Office, October 1, 2003

• Implementing effective security management programs that include consideration of control system security. We have previously reported on the security management practices of leading organizations. Such programs typically consider risk assessment, development of appropriate policies and procedures, employee awareness, and regular security monitoring.

For standardizing purposes, there are several organizations that perform security testing and evaluation of SCADA and control networks. Table 2. presents those organizations.

Sandia National Laboratories has developed several software packages and methodologies, computer aids and etc. in order to find improvement for security of SCADA networks. IDART (Information Design Assurance Red Team) methodology by Sandia National Laboratories is presented by Figure 3.

Table 2.  Organizations to improve control network security

Source: Robert F. Dacey, *Critical Infrastructure Protection*, Testimony Before the Subcommittee on Technology, Information Policy, Intergovernmental Relations, and the Census, House Committee on Government Reform, United States General Accounting Office, October 1, 2003

| Entity | Initiative |
|---|---|
| Sandia National Laboratories | At Sandia's SCADA Security Development Laboratory, industry can test and improve the security of its SCADA architectures, systems, and components. |
| | Sandia also has initiatives under way to advance technologies that strengthen control systems through the use of intrusion detection, encryption/authentication, secure protocols, system and component vulnerability analysis, secure architecture design and analysis, and intelligent self-healing infrastructure technology. |
| Idaho National Engineering and Environmental Laboratory, Sandia National Laboratories, National Energy Technology Laboratory, and other entities | Plans are under way to establish the National SCADA Test Bed, which is expected to become a full-scale infrastructure testing facility that will allow for large-scale testing of SCADA systems before actual exposure to production networks and for testing of new standards and protocols before rolling them out. |
| Los Alamos National Laboratory and Sandia National Laboratories | Los Alamos and Sandia have established a critical infrastructure modeling, simulation, and analysis center known as the National Infrastructure Simulation and Analysis Center. The center provides modeling and simulation capabilities for the analysis of critical infrastructures, including the electricity, oil, and gas sectors. |
| National Science Foundation | The National Science Foundation is considering pursuing cybersecurity research and development options related to the security of control systems. |

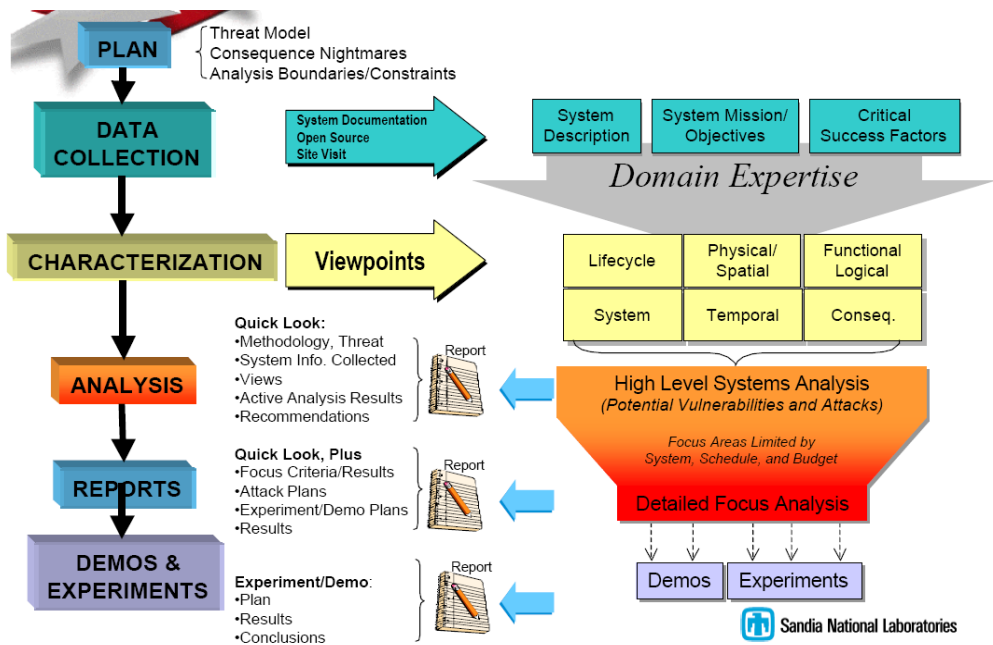

Figure 3.  IDART methodology by Sandia National Laboratories

Source: http://www.naseo.org/committees/energysecurity/energyassurance/hutchinson.pdf

Siemens Company argues[16] that the following are the five steps that could secure a network:

1. Defining the "Security Policy"

A fundamental component of every solution is a corporate "security policy", which defines the significance and organization of IT security for the enterprise. The objectives of any security policy are to create binding specifications and to describe the basic organization. The wholesale support of corporate Management is essential if this kind of policy is to be effectively implemented throughout the company.

2. Security Concept and Security Management

Specific measures are described within the security concept. The concept takes into account the individual business processes and communication links inside and outside the company. In addition, external influences such as legal regulations and requirements imposed by company's business associates are other factors to which consideration has to be given.

The security management necessary for the measures defined in the security concept requires clearly defined responsibilities - in other words, the special security tasks to be performed by each and every staff member have to be clearly defined.

3. Implementing the Security Measures

The measures selected are put into practice during the implementation phase on the basis of the previously defined security policy and the security concept. Implementations of this kind are rarely ready-made, but represent a customized solution in each individual case which takes account of the special requirements of the company. This ensures that the security solution can be harmoniously integrated into the existing infrastructure and the associated system architecture. Nevertheless, it may be necessary, despite this, to adjust individual business and/or communication processes and to redesign them to comply with the security policy. In a few cases, even restrictions or additional effort and cost may be the result.

---

[16] Siemens, *Information Security in Industrial Communications*, October 1999

4. Security Consciousness and Training of Staff Members

The success of all the actions taken towards information security to date is highly dependent upon the attitude of the different staff members. Without the active support of all the persons concerned within the company, it will barely be possible to achieve the specified security objectives. Both decision-makers and all other staff members should therefore be induced to be security-conscious. Suitable means for achieving this are regular security information meetings and specific training sessions.

5. Control and Monitoring

So that a specified security level can be guaranteed permanently, the measures introduced and compliance with the security guidelines have to be monitored and controlled at regular intervals. This is the only way of eliminating existing gaps in security and detecting potential risks.

In conclusion, it is important to point out the uniqueness of the security of web-based industrial networks and SCADA networks. It is obvious, that there are potential threats and risks associated with industrial control networks. The battle for IT security must go on, especially when dealing with security of industrial control networks.